\documentstyle{article}
\font\tenbf=cmbx10
\font\tenrm=cmr10
\font\tenit=cmti10
\font\elevenbf=cmbx10 scaled\magstep 1
\font\elevenrm=cmr10 scaled\magstep 1
\font\elevenit=cmti10 scaled\magstep 1

\renewenvironment{thebibliography}[1]
 { \elevenrm
   \begin{list}{\arabic{enumi}.}
    {\usecounter{enumi} \setlength{\parsep}{0pt}
     \setlength{\itemsep}{3pt} \settowidth{\labelwidth}{#1.}
     \sloppy
    }}{\end{list}}

\begin{document}
\begin{center}{{\tenbf QCD CORRECTIONS TO HIGGS BOSON PRODUCTION
\footnote{Invited talk given at the November, 1992 DPF Meeting}\\}}
\vglue 1.0cm
{\tenrm SALLY DAWSON \\}
\baselineskip=13pt
{\tenit Brookhaven National Laboratory\\}
\baselineskip=12pt
{\tenit Upton, New York~~11973\\}
\vglue 0.8cm
{\tenrm ABSTRACT}
\end{center}
\vglue 0.3cm
{\rightskip=3pc
 \leftskip=3pc
 \tenrm\baselineskip=12pt
 \noindent
We discuss the ${\cal O}(\alpha_s)$
 QCD radiative corrections to Higgs boson production
in the limit in which the top quark is much heavier than the
Higgs boson.  The subleading corrections, of ${\cal O}
(\alpha_s M_H^2/M_{\rm top}^2)$, are presented for
the decay $H\rightarrow \gamma\gamma$ and shown to be small.
\vglue 0.6cm}
{\elevenbf\noindent 1. Introduction}
\vglue 0.2cm
\baselineskip=14pt
\elevenrm

The search for the Higgs  boson of the minimal standard model is one
of the fundamental missions of future high energy colliders such
as the SSC and the LHC.
Over much of the interesting
 Higgs mass range, the dominant production mechanism is gluon fusion.
If the Higgs boson is lighter than $160~GeV$ ($2M_W$), then it will decay
predominantly to $b \overline{b}$ pairs.
Due to the large QCD backgrounds to the $b \overline{b}$
decay mode, it will probably be necessary to
search for the Higgs boson
  in this mass
region through its rare decay modes,
of which the decay $H\rightarrow \gamma\gamma$ has been
discussed the most.  The rates into
modes such as $\gamma \gamma$ are small
and hence knowledge of
 the QCD radiative corrections is essential for assessing the
viability of the signal.

\vglue 0.6cm
{\elevenbf\noindent 2. Gluon Fusion}
\vglue 0.2cm
\baselineskip=14pt
\elevenrm
The basic production mechanism is
gluon fusion through a triangle diagram.
The rate for this reaction
is sensitive
predominantly
to the heavy quarks and has been known for some time.$^1$
 It
is also sensitive to new generations of heavy colored fermions, to the colored
scalars of supersymmetric models and to any other new particles which couple
to gluons and scalars.
Because the Higgs-fermion coupling
of the standard model
 is proportional to the fermion mass,
a heavy fermion does not decouple when its mass is much heavier
than the Higgs boson mass and so this decay provides a window for
observing the consequences of new heavy particles.
In the limit in which $M_{\rm top} >> M_H$, the
rate can be found from the effective Lagrangian,$^2$
$$
{\cal L}={\beta_F\over g_s(1+\delta)}
 {H\over 2 v} G^{a\mu \nu}G_{a\mu\nu}
\quad ,
\eqno(1)
$$
where $\beta_F$ contains only the contribution of heavy fermions to the
QCD beta function.  The factor of $(1+\delta)=1+2\alpha_s/\pi$
 results from the renormalization
of the $H t \overline{t}$ coupling.
For $M_H < 2 M_{\rm top}$, the $M_{\rm top}\rightarrow \infty$
limit is a reasonable approximation to the rate.

The radiative corrections
to the process $g g \rightarrow H X$
can be performed in a straightforward manner
using the effective Lagrangian of Eq. (1)  and the results are
 given in Ref. 3.
Numerically, they are well approximated by
$$
\sigma(gg\rightarrow H)
\sim
 \sigma_0(gg\rightarrow H)\biggl[
1+{\alpha_s \over \pi}\biggl( \pi^2 +{11\over 2}\biggr)\biggr]
\eqno(2)
$$
For $\alpha_s=.2$ this gives an enhancement factor of 1.7.  The results
are of course
also sensitive to the  choice of renormalization scale and
structure functions.

However
 we are not really interested in the region where $M_H<<M_{\rm top}$,
but in the region where they have similar values.  This is the so-called
``intermediate mass'' Higgs boson.  It is thus
 of interest to compute the next
contribution to the series, the terms of
${\cal O}(\alpha_s M_H^2
/M_{\rm top}^2)$ to assess the accuracy of the approximation.
These terms cannot be computed using the
effective Lagrangian, but rather require a direct evaluation of the
relevant  two
loop graphs.  We begin by computing the two-loop graphs contributing to
the amplitude for $H\rightarrow \gamma\gamma$, which form
 a subset of those
required for the computation of $gg\rightarrow H$.
\vglue .6 cm
{\elevenbf{\noindent 3. $H\rightarrow \gamma \gamma$}}
\vglue 0.2cm
\baselineskip=14pt
\elevenrm

We utilize the techniques of
Hoogeveen$^4$
for
evaluating two- loop graphs involving heavy fermions.
This technique has also been successfully used
to compute the $2-$ loop contribution to the $\rho$
parameter from a heavy top quark.$^5$
Each graph gives a result of the form
$$
{\cal A}_i^{\mu \nu}
 =\biggl(  a_i g^{\mu \nu}k_1\cdot k_2+b_i k_1^\nu k_2^\mu
            +c_i k_1^\mu k_2^\nu
\biggr)\qquad,\eqno(3)
$$
where $k_1^\mu$ and $k_2^\nu$ are the external gluon momenta.
Gauge invariance requires that
$$ \sum_i a_i = -\sum_i b_i,
\eqno(4)  $$
where the sum runs over all the diagrams
(the $c_i$ terms do not contribute for on-shell photons).
This serves as a check of our result.

The denominators arising from the heavy-quark propagators  can
be expanded in powers of the external momentum.
For example,
$$
{1\over (q-k_1)^2-M_{\rm top}^2}
={1\over q^2-M_{\rm top}^2}\biggl(1+{2 q\cdot k_1\over
q^2-M_{\rm top}^2}+ ...\biggr)\quad .
\eqno(5)$$
To obtain the terms of ${\cal O}(M_H^2/M_{\rm top}^2)$ each denominator
must be expanded up to terms containing two powers of $k_1$ and
two powers of $k_2$.
After contracting the amplitudes  with
various combinations of $g^{\mu\nu}$ and the external momenta
 and expanding the denominators as in Eq. (5)
all the  contributions have the form
$$
\int {d^n p\over (2\pi)^n}
 \int {d^n q\over (2\pi)^n}
{(p\cdot q, p\cdot k_i,q\cdot k_i, {\rm etc})
\over (q^2
-M_{\rm top}^2)^k (p^2-M_{\rm top}^2)^l (p-q)^2}\quad .
\eqno(6)$$
Using the symmetries of the numerators the relevant
integrals can be reduced to products of one-loop integrals plus
integrals of symmetric form
$$
B_{k,l}\equiv
\int {d^n p\over (2\pi)^n}
 \int {d^n q\over (2\pi)^n}{1\over (q^2
-M_{\rm top}^2)^k (p^2-M_{\rm top}^2)^l (p-q)^2}\quad .
\eqno(7)
$$
 Integrals
of this form are tabulated as a power series in $1/M_{\rm top}^2$ in Ref 4.
and we can directly apply them here.
Our final result
for the fermion contribution to $H\rightarrow \gamma\gamma$
 is then
$${\cal A}_F^{\mu \nu}={\cal A}_F\biggl(g^{\mu \nu}k_1\cdot
k_2-k_1^\nu k_2^\mu\biggr)
$$
with$^6$
$$
{\cal A}_F
=-\biggl({2 \alpha \over
 \pi v} \sum_i Q_i^2\biggr)
 \biggl\{1+{7\over 120} {M_H^2\over M_{\rm top}^2}
-{\alpha_s\over \pi}
\biggl(1-{61\over 270}
{M_H^2\over M_{\rm top}^2}
\biggr)\biggr\} \quad .
\eqno(8)
$$
It is important to note that the QCD corrections are not
just a rescaling of the lowest order result, but
rather have a different dependence on $M_H/M_{\rm top}
$.

The
effect of the radiative corrections
 is shown in Fig. 1 for
$M_{\rm top}=200~GeV$ (solid line)
 and also  for a degenerate doublet of $SU(2)$ quarks
with $M=400~GeV$ (dotted line).
  This figure includes both top and $W$ loops
as can be clearly seen by the threshold at $2M_W$.
The QCD corrections are well under control and are always less
than $4\%$.
Of course, in the standard model, the fermion loop contributions
are dwarfed by the $W$ boson loops which accounts
for the insensitivity of the rate to the QCD corrections.
  In extensions of the
standard model, however, this is not necessarily the case.
In supersymmetric models, for example, it is straightforward to select
parameters for which the fermion loop contribution is significantly
enhanced relative to the $W$ boson loops.

\vglue .6 cm
{\elevenbf\noindent 4. Conclusions}
\vglue 0.2cm
\baselineskip=14pt
\elevenrm

We have computed the ${\cal O}(\alpha_s M_H^2/M_{\rm top}^2)$
contributions to the decay $H\rightarrow \gamma\gamma$
and have found them to be small.  These corrections form a
subset of those required for the
${\cal O}(\alpha_s M_H^2/M_{\rm top}^2)$ computation of the
rate for $gg\rightarrow gH$.  This calculation is in progress.

\vglue .6 cm
{\elevenbf\noindent 5. Acknowledgements}

This work has been done in collaboration with R. Kauffman.
\vglue 0.2cm
{\elevenbf\noindent 6.  References}
\vglue 0.2cm
\baselineskip=14pt
\elevenrm

\vskip 1in
\centerline{\bf Figure Caption}
\vskip .5in
{\bf Fig. 1.}  Ratio of the ${\cal O}(\alpha_s M_H^2/M_{\rm top}^2)$
result for the decay width for
 $H\rightarrow \gamma \gamma$ to the lowest order $M_{\rm top}
\rightarrow\infty$ limit.

\end{document}